Research paper

# Energy dissipation in composites with hybrid nacre-like helicoidal microstructures

Xin Ying Chan[1], Clarence Chua[1], Sharlene Tan[1], Hortense Le Ferrand[1, 2]*

[1]School of Mechanical and Aerospace Engineering, Nanyang Technological University, 50 Nanyang avenue, Singapore 639798

[2]School of Materials Science and Engineering, Nanyang Technological University, 50 Nanyang avenue, Singapore 639798

*hortense@ntu.edu.sg

**Highlights:**

- Natural materials exhibit complex crack deflection paths that increase their toughness.
- Hybrid microstructures between nacre-like and helicoidal arrangments were created.
- PDMS composites with hybrid microstructures were tested under compression.
- The hybrid microstructures at 90 ° angle had a high ductility at failure while being stiff.

**Abstract**

Natural ceramic composites present complex microstructures that lead to tortuous crack paths and confer them high toughness. Current microreinforced composites do not yet reach the same level of complexity in their microstructures, resulting in poorer properties. To achieve complex microstructuration, magnetically-assisted slip casting (MASC) was conducted using a setup with 4 degrees of freedom. Among all possible microstructures, a hybrid design between nacre-like and helicoidal arrangements was selected due to its ability to tilt and twist the crack path. The hybrid microstructured specimen fabricated, consisting of aluminum oxide micro platelets in a silicone matrix, were tested under compression and their mechanical performance compared. Although nacre-like composites exhibited the highest strength and toughness, helicoidal hybrids could show some ductility and higher stiffness. The fabrication strategy proposed here could thus be a simple route to study more complex microstructures in view of increasing the toughness of microplatelet reinforced composites.

1| Introduction

Tortuous crack paths are observed in numerous tough natural materials, such as in the spicules of the venus basket [1], the dactyl clubs of mantis shrimps [2,3], the nacreous layers of seashells [4–6], and teeth [7], for example. Tortuosity arises from deflections at the interfaces between the organic matrix and stiff anisotropic reinforcements that channel the crack into specific directions. Crack deflection and crack twisting under loading then contribute to dissipate energy by the opening of two interfaces and mixing several modes of fracture [8–10]. In natural stiff and strong materials, crack tortuosity is thus one of the key mechanisms to energy dissipation and increase their toughness and damage resistance [11]. Overall, natural materials have microstructures featuring several recurring structural elements among which layered structures and helicoidal arrangements present highly tortuous crack paths associated with high energy dissipation [12,13]. In layered materials such as nacre, 2D reinforcements in the form of microplatelets are arranged in a brick and mortar fashion. This configuration allows for spreading the load over a larger area. Also, as microscopic cracks get deflected at the microplatelet-matrix interface repeatedly, a staircase crack path is formed, resulting in a macrocrack deflected at an angle $\alpha \sim 58 \pm 7°$ with respect to the horizontal plane [4,5] (**Figure 1A**). In helicoidal arrangements such as in the dactyl club of shrimps, 1D fibrous inorganic reinforcements are stacked and skewed at an angle $\gamma$ to form a helix. The pitch angle $\gamma$ varies from species to species, with values of 45-90° in the scales of *Arapaima gigas* [14], 1-6° in the dactyl club of mantis shrimps [2,8] and 12-18° in beetles' exoskeletons [15,16]. The crack path differs greatly from the brick and mortar structures as it twists along the microstructure, leading to vertical and horizontal cracks in the cross-section (**Figure 1B**) [3,17]. Macroscopically, the damage is extended over a large volume with horizontal cracks at $\alpha \sim 7 \pm 4°$ bridging vertical cracks ($\alpha \sim 98 \pm 10°$) [18]. Microstructures with helicoidal arrangements are found to exhibit a higher toughness and resilience as compared to nacre and other natural materials with similar stiffness and hardness [19]. Indeed, the dactyl club has a strain to failure of 37% for a hardness of 2 GPa whereas nacre has a strain to failure of 1% only for similar hardness [20,21].

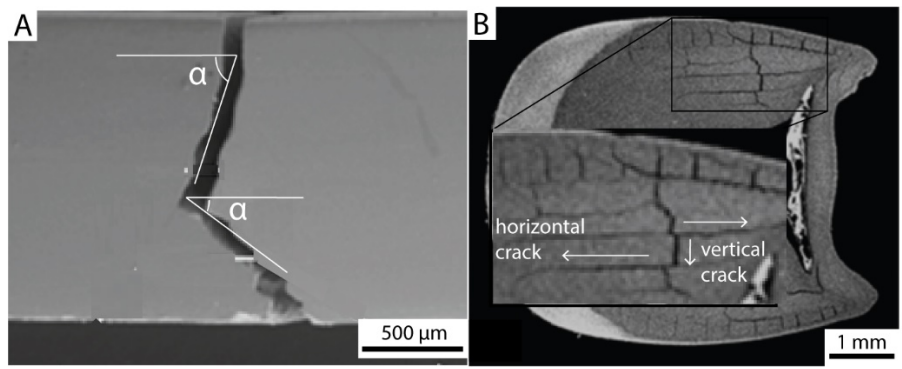

**Figure 1: (A)** Crack propagating in sheet nacre. Reproduced with permissions from Ref [4]. Copyrights 2018, Springer Nature. **(B)** Crack propagating in the periodic region of the dactyl club

of the mantis shrimp. Reproduced with permissions from Ref [18]. Copyrights 2019, National Academy of Sciences.

Although both nacre and helicoidal structures exhibit tortuous crack paths through crack deflections, helicoidal arrangements present also a large amount of crack twisting and mixed modes of crack opening [8,9]. With the abstraction of the other toughening mechanisms such as crack bridging, platelet pull-out and intramolecular interactions, it has been postulated that helicoidal structures are more performant than nacre under multiaxial loading [22]. Helicoidal microstructures also seem to be tougher as shown by the ability of shrimps to break seashells covered with nacre [23]. Building a helicoidal microstructure using microplatelets instead of fibers could thus possibly increase the toughness of nacre-like structures by additional energy dissipation through crack twisting. To date, there are many examples of nacre-like and helicoidal composites, that show high mechanical performance [24–28]. However, no fabrication method controls the pitch angle of a hybrid design coupling nacre-like brick-and-mortar and helicoidal arrangements, *i.e.* locally tuning the microplatelet orientation in 3D, and this in specimens with high solid loadings. Therefore, these microstructures have not yet been explored.

Brittle and highly concentrated microstructured materials with controlled microplatelets orientations have been realized using external electric and magnetic fields, ultrasounds, and shear forces [29]. Among these methods, external magnetic fields are the most used for ceramic-polymer composites. While rotating magnetic fields can orient 2D magnetically-responsive platelets along a specific plane [30], consolidating the structure and maintaining the orientation can be realized by polymer curing or gelling [31] or slip casting [32,33]. On the one hand, polymer curing or gelling are interesting as they can be combined with 3D printing and multiple local orientations. However, the microplatelet concentration is typically low, of 5 to 10 vol% [34,35]. Slip casting, on the other hand, yields brittle composites at solid loadings of ~40 vol%, depending on the microplatelet's dimensions and the slurry concentration [32]. Furthermore, magnetically-assisted slip casting (MASC) with periodic microstructures and controlled microplatelet orientation could be achieved [36]. These microstructures presented layers with microplatelet orientation control of one angle at a time, which were the rotations around the horizontal axis or the vertical axis [36,37]. Crack tortuosity has been reported in ceramics with aligned orientations of platelets and densified by pressureless sintering [38], as well as at the junction between layers with perpendicular orientations [36]. However, crack tortuosity in complex microstructures has not been looked at in more complex microstructures due to the lack of an automated and controllable processing method. To be able to create composites with orientation control of microplatelets in 3D using magnetically-assisted slip casting, there is thus the need for a new setup. Furthermore, the energy dissipation and toughening of such complex microstructures have not yet been studied, except for the nacre-like microstructures.

Since crack deflection is one of nature's tricks to increase energy dissipation in composites, this paper aims at using this strategy to increase the dissipation in ceramic microplatelets reinforced materials. First, a setup is developed to allow the tuning of platelet orientation in the 3 dimensions of space during MASC. Although the principle of MASC has been demonstrated in earlier work, no set-up was yet built and tested to allow for the controlled orientation of microplatelets in the 3 directions of space. The resolution control and degree of freedom achieved to create complex microstructures are described. Then, polymer composites with hybrid microstructure were prepared and tested under quasi-static compression tests. Their crack path and stress-strain curves were measured and compared to better understand the microstructure-properties relationships. Increasing the energy dissipation of materials by microstructure control has the potential to increase the toughness of ceramics and composites reinforced with 2D particles such as ceramic-graphene, ceramic-metal or ceramic-polymer composites, and to create 3D structures that sustain loadings applied in multiple directions.

## 2| Materials and Methods

*Slurry preparation and characterization.*

Alumina microplatelets ($Al_2O_3$, Ronaflair, Merck), of average diameter 8.5 $\mu m$ and thickness 300 nm, were magnetized following a method described previously [32]. In short, 0.01 vol% of iron oxide nanoparticles (EMG-705, 10 nm diameter, Ferrotec), were mixed with the $Al_2O_3$ microplatelets in water at neutral pH until complete electrostatic adsorption was achieved. The resulting magnetic platelets, m-$Al_2O_3$, were then filtered out and rinsed with ethanol and water before being thoroughly dried. These functionalized particles were then used for the slurry. The slurry consisted of 20 vol% m-$Al_2O_3$ dispersed in water in presence of a few drops of Dolapix CE64 (Zschimeer & Schwarz, Germany). After ultrasonication (Sonopuls HD 4100, Bandelin, Germany), 5 times 2 minutes each at 20% amplitude, 5 wt% of poly(vinyl pyrrolidone) (PVP, MW=360'000, Sigma-Aldrich, China) was added to the suspension and dissolved and mixed for a few hours. The rheological properties of the slurry were characterized using a rheometer (Bohlin, Malvern Instruments, UK) with serrated parallel plate set-up under controlled shear rate at room temperature. Viscometry was carried out with a shear rate ramp of 0.01 $s^{-1}$ to 100 $s^{-1}$. Oscillatory rheology was performed under frequency of 1 Hz with amplitude sweep from 0.1 Pa to 1000 Pa and delay time of 2 s.

*Slip casting.*

Porous gypsum substrates were prepared beforehand using the recommendations from the manufacturer (Ceramix, Germany). The substrates were kept in ambient conditions of temperature and humidity (25 °C and 80% humidity). The same substrates were used throughout the study. The molds for the casting were prepared by adding a plastic tube on top of the gypsum. The casting kinetics was measured by dipping a plastic rod after casting the slurry to measure the remaining thickness of the liquid.

*Magnetic alignment setup.*

The magnetic alignment setup was built using 3D printed parts (Prusa I3 MK3S MMU2S, BQ 1.75 mm PLA filament), servo-controllers (SKU 900-00008 and SKU 900-00005, Parallax Inc, Continuous rotation servo and standard servo), and an Arduino interface. The motors were powered by a DC controller (MS 3010D, Maisheng, China) operating at 5V. The magnet for the magnetic orientation was a Neodymium magnet of magnetic strength 150 mT at the surface and rotating at 1 Hz. The sample was positioned close to the magnet where the magnetic field strength is 14 to 60 mT. The magnetic field strength was measured using a Gaussmeter (Hirst Magnetics GM07, RS Pro, Singapore).

*Sample orientations characterization.*

After casting and aligning using our magnetic alignment set-up, the samples were left to dry for 48 hours at 48 °C (IKA oven 125 control dry, Malaysia). The samples were then carefully unmolded by using a spatula to gently push on one side of the sample until it was freed from the mold. The alignment patterns at the surface of the samples were observed using an optical microscope (Dino-lite AM7915MZTL, Labfriend, Singapore, 10-140 times magnification) placed at 90° from the thickness of the samples (see Supplementary Information (SI) **Figure S1**). The images were analyzed using Image J (NIH, USA). Surfaces obtained by brittle fracture of the samples were imaged using a scanning electron microscope (JSM-5600LV SEM, JEOL Asia Pte Ltd, Singapore) after 120 s sputtering with gold.

*Infiltration of the scaffolds.*

After drying, the porous scaffolds were infiltrated with a mixture of polydimethylsiloxane (PDMS) (SYLGARD™ 184 Silicone Elastomer, Dow Corning Corporation, USA) and low viscosity silicone oil (10 cSt Silicon oil, Sigma-Aldrich, Singapore). First, PDMS was mixed with its curing agent and silicone oil in the ratio 10:1:11. The mixture was stirred for 2 minutes and degassed in a vacuum for 5 minutes. The liquid solution was then cast on the scaffold and placed in a vacuum for infiltration (Binder VD 53, Fischer Scientific Pte Ltd, Singapore) until no more bubbles were observed. The samples were removed from the solution and left to cure at 48 °C for 72 hours (IKA oven 125 control dry, Malaysia). Excess PDMS surrounding the composites was then peeled off. The infiltration did not impact the microstructure (see SI **Figure S2**). The overall process is summarized in **Figure 2.** After curing the polymer matrix, the samples were polished into cylinders for mechanical compression testing.

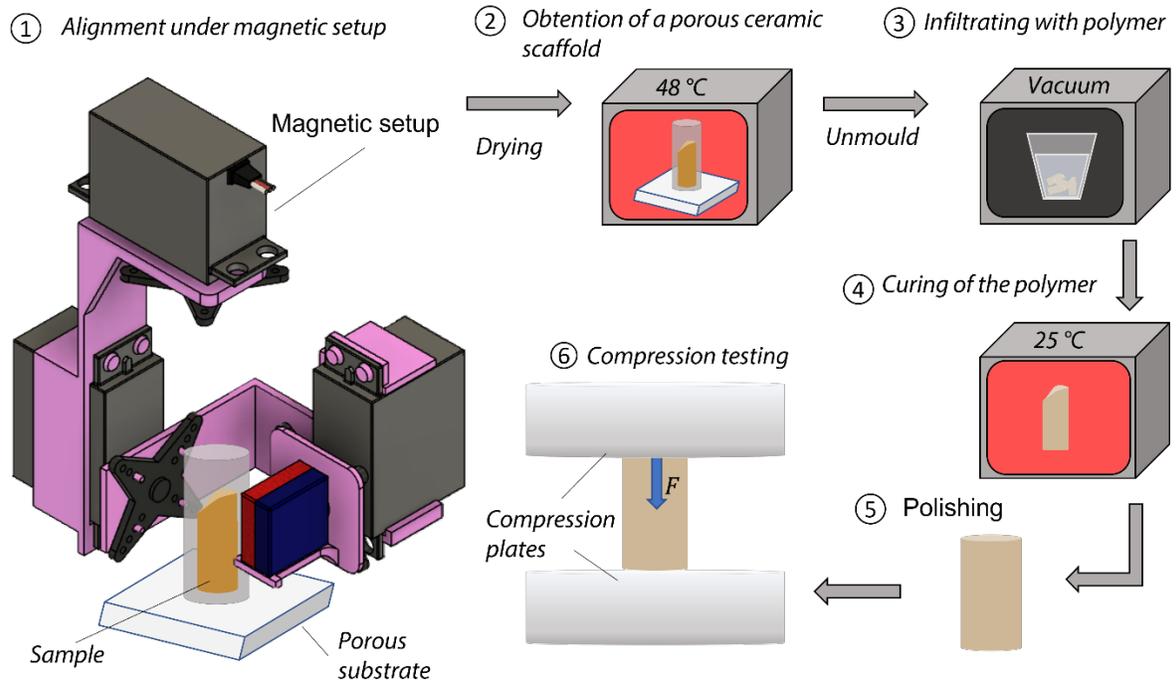

**Figure 2:** Schematics of the processing steps for the fabrication of microstructured composites for compression testing.

*Mechanical testing.*

Compression testing was done on a Universal testing system, Instron 3366, with a 500 N load cell and at a strain rate of 1%, until fracture. 5 samples of each type at least were fabricated and tested. Videos were recorded during the compression using the optical microscope at 90 ° to the thickness of the samples (Dinoscope). The mechanical values measured and calculated from the compression test are represented in **Figure 3**. The compressive load $F$, displacement $\Delta h$ from the universal tester, and sample height $h$ and diameter $d$ were used to calculate the engineering stress $\sigma$ and strain $\varepsilon$ using the following formulas

$$\sigma = \frac{4F}{\pi d^2} \quad \text{(eq 1)}$$

$$\varepsilon = \frac{\Delta h}{h}. \quad \text{(eq 2)}$$

The energies absorbed elastically and dissipated through cracking were calculated by applying the midpoint rule to the data points. The pre-crack propagation energy $E_{pre-crack}$, comprising elastic strain energy and plastic work done by crack initiation was calculated as the area under the engineering stress-strain curve up until the ultimate compressive stress $\sigma_{max}$, following:

$$E_{pre-crack} = \int_0^{\varepsilon_{max}} \sigma \, d\varepsilon \quad \text{(eq 3)}$$

The total energy dissipated by the sample $E_{diss}$ was calculated until the strain $\varepsilon_{fail}$, at which the stress has decreased significantly to a point corresponding to 20% of $\sigma_{max}$ and a macroscopic crack appeared. The energy dissipated by cracking $E_{crack}$ was then calculated as

$$E_{crack} = E_{diss} - E_{pre-crack} \quad \text{(eq 4)}$$

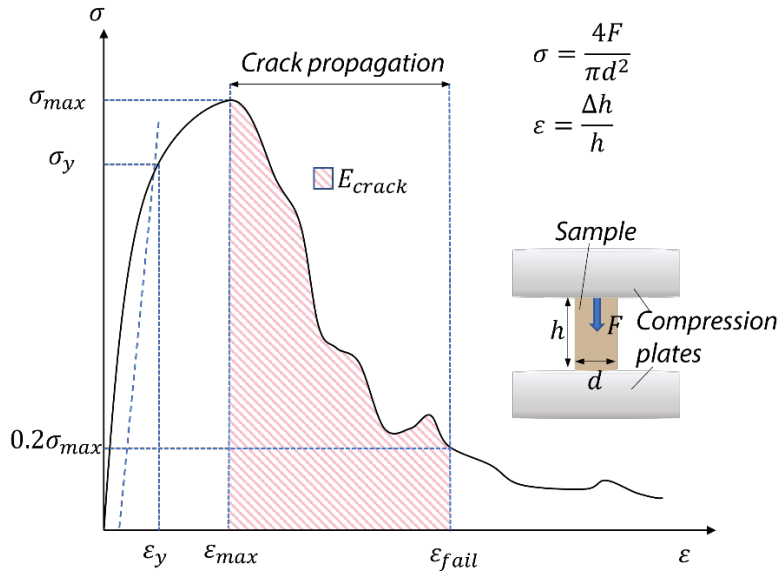

**Figure 3:** Schematics of the mechanical compression test and data analysis.

## 3| Manufacturing principle, set-up and microstructures

### 3.1| Slurry preparation

The fabrication approach for our microstructured samples follows the magnetically-assisted slip casting method developed previously [36]. In short, anisotropic alumina platelets are rendered magnetically responsive by surface decoration with iron oxide superparamagnetic nanoparticles (**Figure 4A**). Such particles orient under low magnetic fields of strength > 3 mT and rotate typically at frequencies above 1 Hz. Depending on the viscosity of the fluid and the magnetic field strength, the frequency required for the alignment can be tuned accordingly following an established relation [30]. For slip casting, a slurry is prepared that contains 20 vol% of magnetic platelets (m-$Al_2O_3$) dispersed with a surfactant and a small concentration of polymer as a binder. At this concentration, the slurry presents shear-thinning properties with a flow index $n = 0.759$ and a consistency index $K = 0.995$. $n$ and $K$ are defined following the Herschel-Bulkley model(**Figure 4B**):

$$\tau = \tau_0 + K \cdot \dot{\gamma}^n, \tag{eq 3}$$

where $\tau$ is the shear stress, $\tau_0$ a constant and $\dot{\gamma}$ the shear rate. The slurry also presents a liquid-like behavior with $G' < G''$, $G'$ and $G''$ being the elastic and storage modulus, respectively (**Figure 4C**). Liquid-like properties and flowability of the slurry are important to enable casting and magnetic alignment, whereas the concentration of particles enables locking of the particle's position after casting and solvent drainage through the porous substrate [32]. This casting occurs with a deposit thickness $x$ increasing with the time $t$ following:

$$x = A \cdot \sqrt{t} \qquad \text{(eq 4)}$$

with $A$ being a constant measured experimentally (**Figure 4D**). For our system, $A = 48 \pm 0.34$ s.mm$^{-2}$. The casting kinetics plays an important role in controlling the pitch $p$ of the layered structures made thereafter and is dependent on environmental conditions such as temperature and humidity, slurry composition and substrate material. Although similar slurries have been cast in other works, the different settings required us to characterize the slurry viscosity and casting kinetics again.

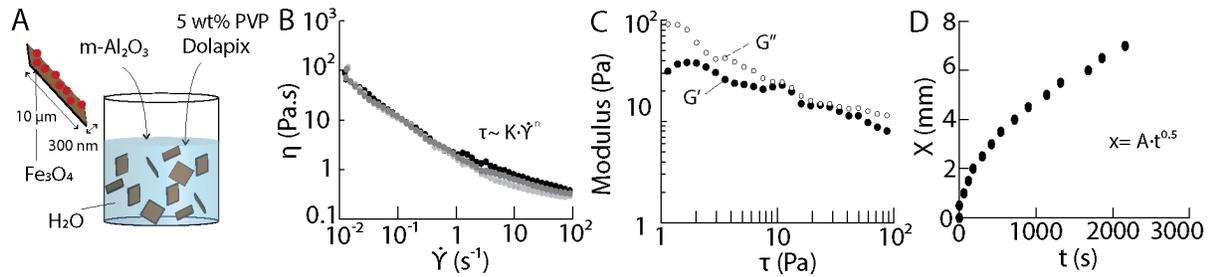

**Figure 4: Slurry preparation and casting kinetics.** **(A)** Illustration of the composition of the slurry. PVP stands for poly(vinyl pyrrolidone) and m-Al$_2$O$_3$ for the magnetized alumina microplatelets. **(B)** Viscosity $\eta$ as a function of the shear rate $\dot{\gamma}$ for two slurries of identical compositions. $\tau$ is the shear stress, $K$ the consistency and $n$ the flow index. **(C)** Elastic (G') and viscous (G'') moduli as a function of the shear stress $\tau$ for the slurry. **(D)** Deposit thickness $x$ as a function of the casting time $t$. $A$ is the constant representing the casting kinetics.

### 3.2| Fabrication setup for 3D platelet orientation and controlled layer thickness

To orient the platelets in the 3 dimensions of space, 2 angles, defined as $\theta$ and $\phi$ (**Figure 5A**, insert) need to be controlled simultaneously. In other works, only one of these two angles had been varied [32,36,37]. The difficulty to achieve both angles is to have a set-up that is stable, programmable, and controllable. For this purpose, we built a setup that features 3 motors. Motor 1 supports a magnet and spins continuously at a frequency above 1 Hz to orient the microplatelets in the plane of rotation of the magnet, called the plane of alignment. This plane of alignment is then turned by 2 other motors that move in a step-wise fashion as controlled by the computer interface: motor 2 tilts the plane by an angle $\theta$, whereas motor 3 tilts the plane by an angle $\phi$. By moving the motors 2 and 3 at defined repetitive angles with a set timestep of $t_0$ during casting, layers with preprogrammed angles were built. The layers were visible by a change of color due to the reflectivity of the plane of the platelets (**Figure 5B,** see Supplementary **Figure S3** for the variation in reflectivity with the platelet angle). With this setup, it is now possible to create various microplatelets assemblies (see **SI movie S1 and S2**) with platelets tilted at any desired angle (**Figure 5C**).

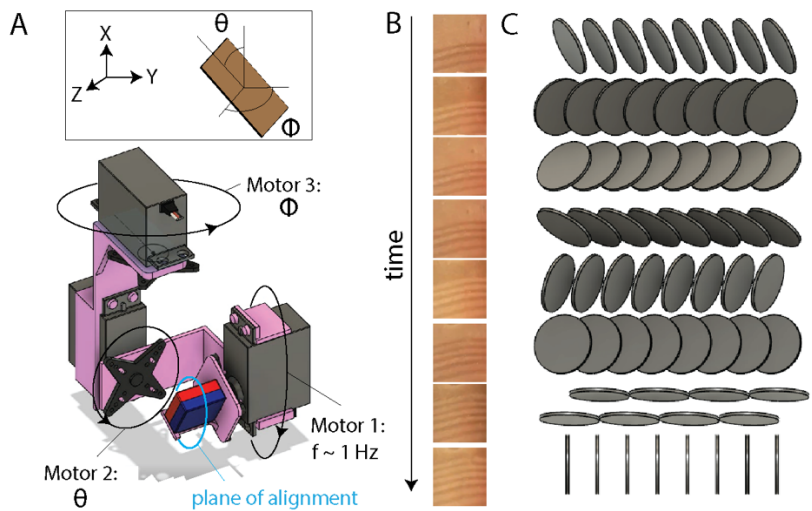

**Figure 5: Fabrication setup. (A)** Schematics depicting the magnetic assembly set-up and the angular orientations of the platelets. Motor 1 controls the rotation frequency of the magnet, Motor 2 controls the angle $\theta$ of the platelet and motor 3 the angle $\phi$. Insert shows the definition of these angles in the global coordinate system (X, Y, Z). **(B)** Build up of oriented layers with time. **(C)** Schematics of a microstructure with varying $\theta$ and $\phi$.

### 3.3| Microstructure control and limits

To define the limits of the system's capabilities, we first measured the minimum timestep $t_0$ that can be used. For this, we chose a variation of angle $\theta = 90°$ while $\phi$ remained constant (**Figure 6A**). The smallest $t_0$ obtained was $t_0 = 5$ s, which sets the smallest layer thickness attainable. This time corresponds to the minimum time required for the alignment of platelets in the slurry [32]. Below 5 s, no platelet orientation could be found in the sample. The measurement of the pitch followed the casting kinetics with pitch variation described by [36]:

$$p = \frac{A^2 \cdot t_0}{2x} \tag{eq 5}$$

Although the pitch varied with the thickness of the samples when the timestep was kept constant, layers of less than 100 µm could be created with layer thicknesses as low as 10 µm for the horizontal alignment of the platelets and 20 µm for the vertical orientation above $x = 4$ mm (**Figure 6B,C**). Next, we tested our setup for various tilt angles $\theta$ at constant time step $t_0 = 30$ s (**Figure 6D**). The patterns created followed the trend although the fit of the pitch had some deviations from the prediction for $\theta = 30°$ and $\theta = 60°$ (**Figure 6E**). Presumably, the casting kinetics varied with the tilted angles but could not be measured experimentally. Nevertheless, looking at the color patterns that developed along with one pitch, several plateaus could be recognized that matched with the number of layers, with 6 angles for a variation in $\theta$ of 30°, 3 angles for a variation of 60°, and 2 angles for a variation of 90° (**Figure 6F**). The sample with no variation in angle did not exhibit any color change. This particular sample had a different color than the other samples (**Figure 6D**) because the platelets were prepared with a new batch of iron oxide nanoparticles, but exhibited a similar response. As the motor step control had a precision of 1°, any other angle step could be controlled using this setup. However, the misalignment obtained with magnetically-aligned samples is typically around 10° and increases to ~20° as the verticality of the platelets increases [32,33,39]. Finally, to demonstrate the feasibility of changing simultaneously the angles $\theta$ and $\phi$, we constructed microstructures with helicoidal twist and tilt patterns (**Figure 7**). The SEM images reflect the helicoidal tuning of the platelets.

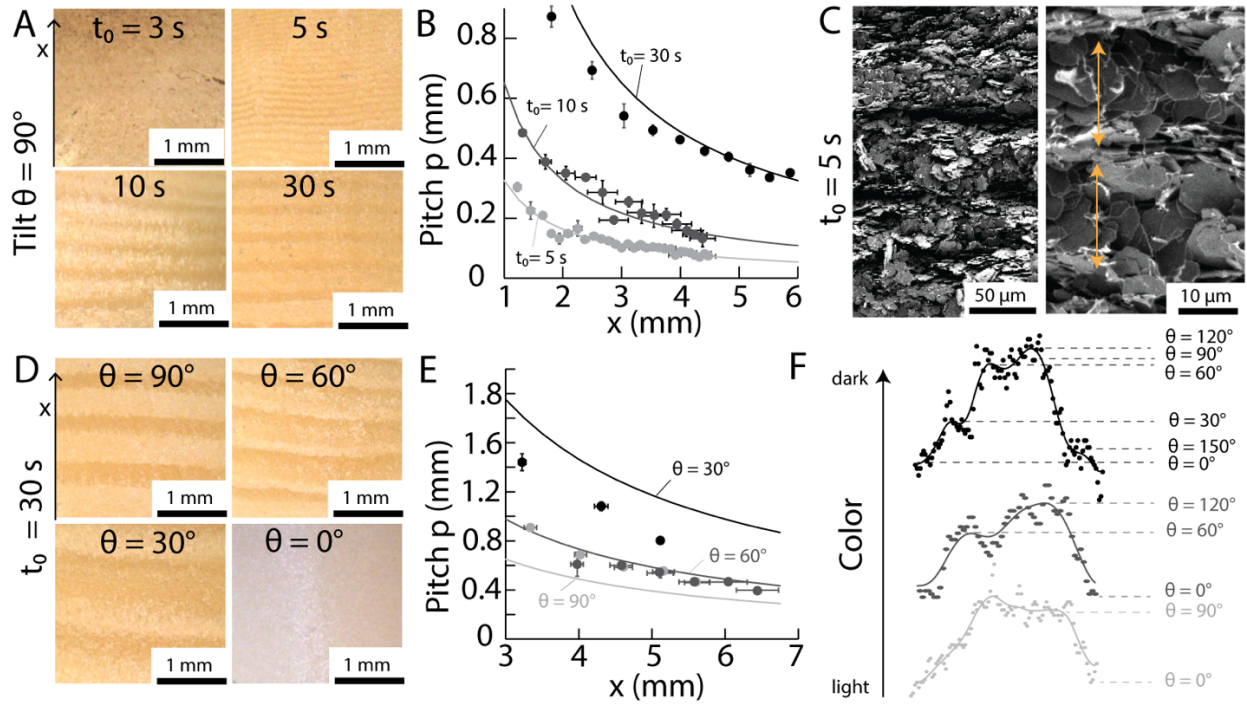

**Figure 6:** (A) Optical images of the patterns in samples with $\theta = 90°$ and varying timestep $t_0$ between the change in $\theta$. (B) Pitch $p$ of the patterns with $\theta = 90°$ as a function of the sample thickness $x$ for varying timestep $t_0$. (C) Close-up electron micrographs showing the fine microstructure for $\theta = 90°$ and $t_0 = 5$ s. (D) Optical images of the patterns in samples with $t_0 = 30$ s and varying $\theta$. (E) Pitch $p$ of the patterns with $t_0 = 30$ s and varying $\theta$ as a function of the sample thickness $x$. **(F)** Variation of color within one pitch indicating the presence of 2, 3 and 6 layers for $\theta = 90, 60$, and $30°$, respectively.

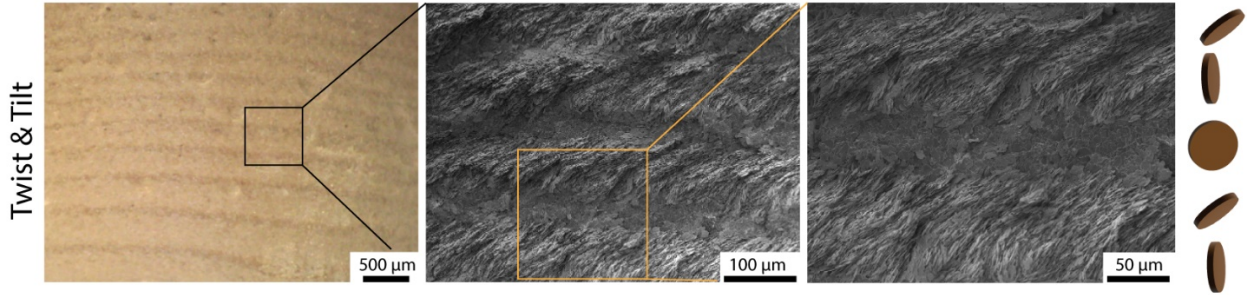

**Figure 7:** Optical image and electron micrographs of a helix structure with simultaneous variations of twisting and tilting angles $\theta$ and $\phi$. The schematic on the right indicates the orientation of the microplatelets within each layer.

**4| Results**

**4.1| Samples microstructures and cracking under compression**

To study the effect of microstructure on the mechanical properties of the composites under compression, a set of 8 microstructures was produced **(Figure 8)**. The porous scaffolds were infiltrated with a mixture of PDMS and silicon oil at 50-50 vol%. Although this matrix is remarkably tough (see **SI Figure S4**), the samples remained brittle. Scanning electron micrographs showing the microstructure of the specimen are presented in the supplementary materials (**SI Figures S5,6,7**).

Monolithic composites comprising only one constant platelet angle at 0, 90 and 45° with respect to the vertical were first made and tested (**Figure 8A**). These samples have a microstructure similar to the nacre-like microstructure. Then, we produced multilayered composites with varying angles between each layer and using $t_0 = 30\ s$ (**Figure 8B**). These samples are what we refer to as hybrid nacre-like helicoidal microstructures. Before looking at the stress-strain curves, we looked at the path of the cracks during fracture. The images in **Figure 8** are representative of the cracking of the samples tested. All samples cracked during the compression test. In monolithic microstructures, the cracks were straight and along the platelet's plane orientations for samples with $\theta = 0°$ (vertical orientation) and $\theta = 45°$ (tilted orientation). This behavior is similar to what has been reported elsewhere [40]. The samples with $\theta = 90°$ (horizontal microplatelet orientation) showed a combination of vertical and tilted microcracking, which is also expected from horizontal nacre-like microstructures [39]. Multilayered specimens showed different cracking paths with crack bifurcation and multiple microcracking. No visible significative difference could be observed in the cracking path between the multilayered samples.

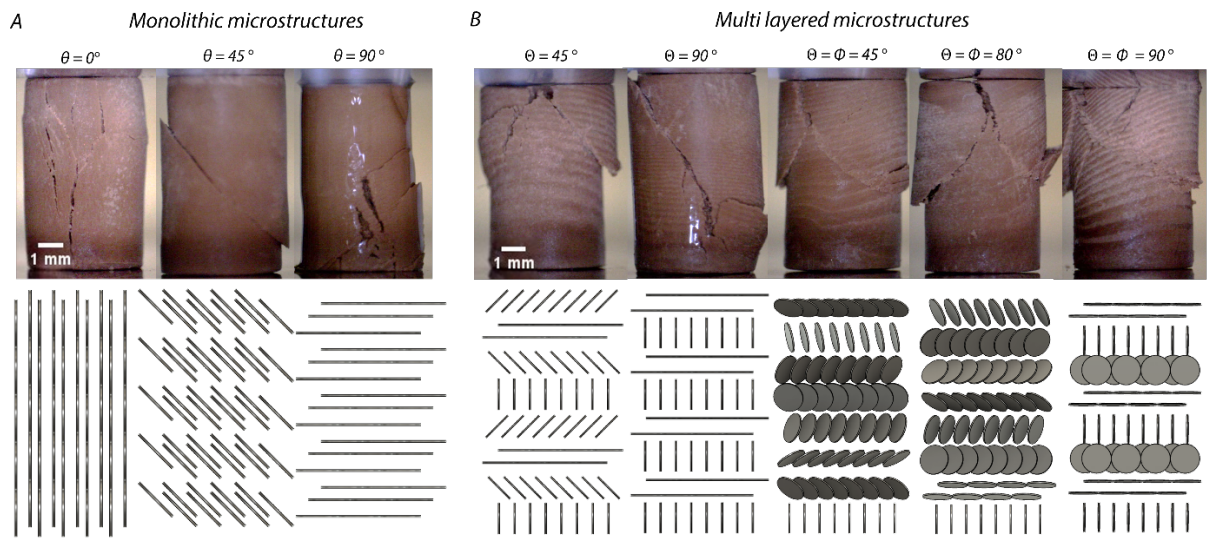

**Figure 8:** Optical images of the microstructured composites produced and tested in this study, with monolithic microstructures **(A)** and multilayered microstructures **(B).** The pictures were taken during the compression test to show the crack paths in the samples. The schematics below each sample indicate the microplatelets orientations within the structures. The angles $\theta$ are indicated for the monolithic samples, whereas the angle steps between each layer are indicated for the multilayered microstructures ($\theta$ and $\Phi$). The scale bar on the optical images is the same for all pictures.

**4.2| Stress-strain curves of the microstructured composites under compression**

We tested our microstructured composites under compression at a constant strain rate of 1%/s and recorded their stress-strain curves (**Figure 9**). The microstructure had a significant effect on the mechanical properties of the samples. Although there are some variations in the measured properties, probably due to non-perfect surface finishing due to the brittle nature of the composites, the results represented in **Figure 9** are representative of the measurements and are reproducible.

Monolithic samples showed higher yield stress, strength, and maximum strain as the angle $\theta$ increased from 0 to 90° (**Figure 9A-C**). Usually, microstructured composites are stronger along the direction of alignment of their filler. Therefore, it may appear that our results contradict this statement since our samples are stronger when they are loaded perpendicularly to the filler orientation. However, most literature report tensile tests instead of compression. In vertical and unconfined compression as used here, tensile stresses develop horizontally, perpendicularly to the loading direction. Furthermore, the matrix used in these composites, PDMS, is strong in compression but weak in tension, with a maximum tensile strain of less than 2% [41]. Therefore, it is consistent that the samples with horizontally aligned microplatelets ($\theta = 90°$, **Figure 9C**) exhibited the highest strength, yield stress, and strain at fracture. Indeed, the matrix can take up most of the compressive load while being reinforced along the directions of the tensile stresses.

The large drop in the stress at about 7 % strain occurs when there is a bifurcation in the crack path which suddenly goes from vertical to horizontal, along the microplatelet plane. For the samples that have vertically aligned microplatelets ($\theta = 0°$), there is no reinforcement against the tensile stresses developing horizontally and crack occurs readily. For the intermediate angle ($\theta = 45°$), the crack path is oriented at 45° along the platelet direction, indicating very weak reinforcement as well and properties similar to that of $\theta = 0°$.

The multilayered microstructured specimens also exhibit mechanical properties varying with the angles of the microplatelets (**Figure 9D-H**). All microstructures, however, had lower yield stress and strain as compared to the monolithic horizontal nacre-like $\theta = 90°$ samples. This could be due to lack of reinforcement along the tensile stresses directions, but also the occurrence of delamination at the interface of the layers. In general, the microstructures containing layers with horizontally aligned microplatelets had higher mechanical properties than the other ones. In these sets of microstructures, the layer thickness was not constant but decreased from the top to the bottom. We noticed that the thinner layers were more prone to cracking than the thicker layers. The effect of the layer thickness on the mechanical properties and cracking will be studied in a follow-up study.

Based on these stress-strain curves, the mechanical data and energy dissipation potential of the composites were measured and compared, as discussed in the following section.

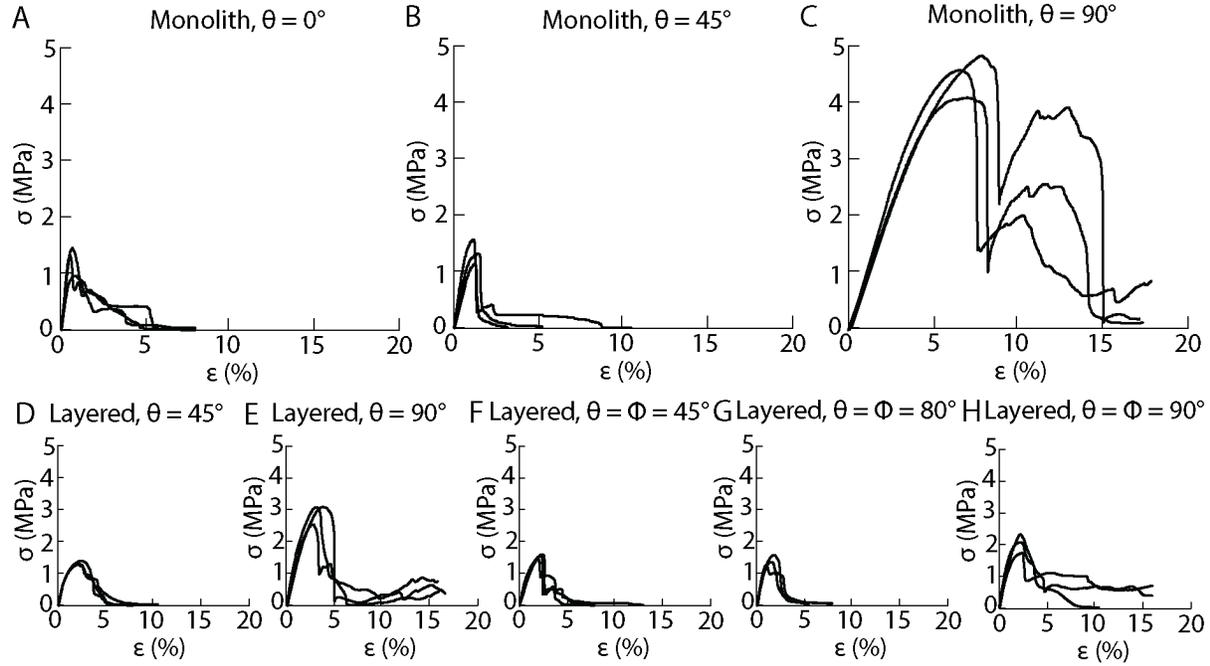

**Figure 9:** Stress-strain curves under compression of the monolithic microstructured samples with microplatelets angle $\theta = 0°$ **(A)**, 45° **(B)** and 90° **(C)** and of the layered microstructured samples with angles steps $\theta = 45°$ **(D)**, $\theta = 90°$ **(E)** and $\theta = \phi = 45°$ **(F)**, $\theta = \phi = 80°$ **(G),** and $\theta = \phi = 90°$ **(H),** as illustrated in **Figure 8.**

**4.3| Microstructure-property relationships**

We compared the Young's moduli, stresses at yield, maximum and failure, the corresponding strains at these stresses, and the energy dissipated before and during crack propagation of the different microstructured composites (**Figure 10**).

First, the Young's moduli of the monoliths showed significant decrease from the vertically aligned samples to the horizontally aligned, from ~280 MPa to ~92 MPa, respectively (**Figure 10A**). However, the Young's moduli of the multilayered microstructures did not vary significantly with the angles. The Young's moduli of these samples were about 140-160 MPa, which is around the average of the moduli of the vertical and horizontal monoliths. Using a filler concentration of 35% and Young's moduli of 0.1323 MPa and 330 GPa for the matrix and the microplatelets, respectively, the upper and lower bound moduli from the Voigt and Reuss models were estimated at 116 GPa and 0.2 MPa, respectively. The stiffness of our composites is thus closer to the lower bound than the upper bound, which is likely due to the micrometric dimensions of the fillers and their weak bond to the matrix. Sintering the microstructures before infiltration would likely create much stronger composites by allowing diffusion and bonding between the ceramic particles.

Then, the stresses, strains, and energies dissipated of the composites also varied significantly with the microstructure (**Figure 10B,C,D**). Overall, the monoliths with horizontally-aligned microplatelets performed better for all these properties. The multilayered sample with $\theta = \phi = 90°$ also exhibited a high ductility at failure similar to the horizontally aligned monolith, which lead to higher total energy dissipated as compared to the other multilayers. The strain at the yield point for this sample with $\theta = \phi = 90°$ was nevertheless lower. The energy dissipation of the composites was divided into two contributions: the contribution from the energy dissipated elastically and plastically before cracking, $E_{pre-crack}$, and the contribution from the energy dissipated during microcracking, $E_{crack}$. The relative contributions of those two energies were similar for all samples except for the vertically aligned monoliths and for the multilayer composites with $\theta = \phi = 90°$ for which the energy dissipated by microcracking was superior to the elastic energy. This is likely due to the vertically oriented layer that cracked easily under the vertical compression loadings.

Finally, we plotted a radar chart to better visualize and compare the overall performance of the composites, namely their stiffness, strength, ductility, and toughness (**Figure 10E**). It appears that the monoliths with horizontal, nacre-like orientation ($\theta = 90°$) display significantly higher strength and toughness as compared to the others, but with lower stiffness. This result is expected since the composite has a nacre-like microstructure which is known for a superior combination of strength and toughness [39]. The second-best performing microstructure is the multilayered $\theta = \phi = 90°$ which has a combination of vertical and horizontal layers. However, this microstructure still exhibited a toughness 4 times less than the nacre-like monoliths. Since toughness can only be obtained with the horizontally aligned layers, the layer thicknesses in the multilayers at $\theta = \phi = 90°$ could be varied to further increase the strength and toughness, and the stiffness. Also, despite the other microstructures being less performant under the vertical compressive loading, they could be tested under other loadings directions, where some layers have the plane of the microplatelets oriented parallel to the load. Micro-structuring could therefore be an interesting alternative for samples to sustain loads applied at various directions.

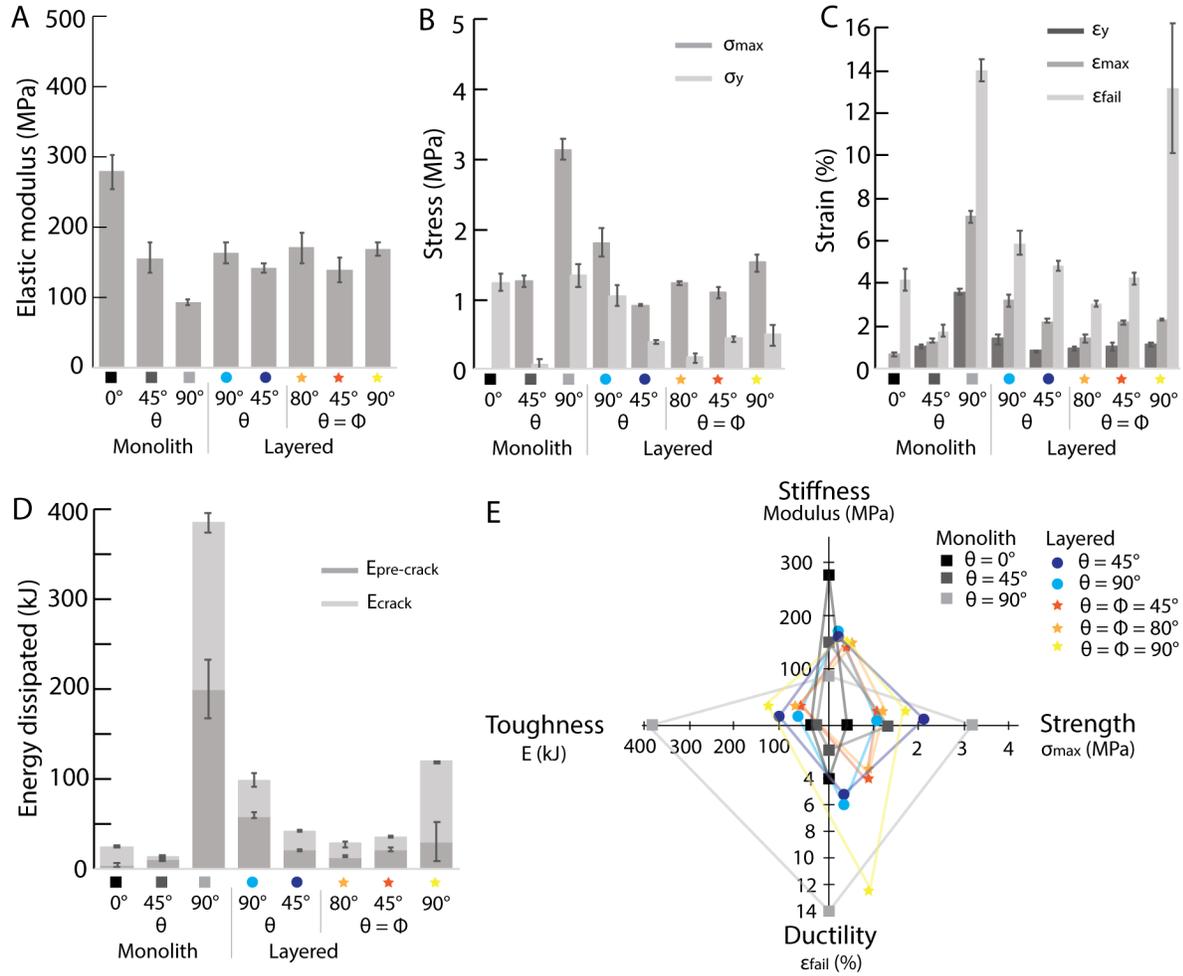

**Figure 10: Mechanical properties of the microstructured composites:** elastic modulus **(A)**, yield $\sigma_y$ and maximum stresses $\sigma_{max}$ **(B)**, strain at yield point $\varepsilon_y$, at the maximum stress $\varepsilon_{max}$, and at failure $\varepsilon_{fail}$ **(C)**, energy dissipated by cracking $E_{pre-crack}$ and during cracking $E_{crack}$ **(D)**. **(E)** Radar plot comparing the stiffness, strength, ductility and toughness of the samples. The toughness represented is the sum of $E_{pre-crack}$ and $E_{crack}$.

## 5| Discussion

The composites explored in this study had PDMS as a matrix. A much stronger and tougher matrix would further enhance the properties. For example, polyether urethane diacrylate-co-poly(2-hydroxyethyl methacrylate), a tough polymeric matrix, has been found to yield higher strength and toughness as harder and brittle and soft and flexible matrices in microstructured ceramic-polymer composites at ceramic concentration 55 % [42]. The reported flexural modulus for these composites was 36.8 GPa for a fracture strength of 168 MPa and a work of fracture of 156.2 J.m$^{-2}$, obtained under three-point bending tests. Another work on alumina microplatelet reinforced composites used a commercial epoxy as a matrix and conducted tensile experiments [31]. The configuration with the long axis of the microplatelets oriented along the direction of the tensile forces also leads to the highest yield stresses and strain, of values ~45 MPa and 5%, respectively. The ductility is much lower than in our composites and the strength 10 times higher, probably due to the stiffness of the epoxy matrix used. Microplatelet composites films with nacre-like orientations in chitosan matrix were also found to exhibit larger ductility up to 25%, and higher tensile stresses between 200 and 300 MPa [43]. However, these films contained only 10 to 15 vol% of microplatelets, whereas our composites had a solid loading of about 35-40 vol%. These results nevertheless indicate that the combination of microstructure control, as proposed and enabled in this study, with a carefully chosen matrix could optimize the mechanical performance. Finally, another study looked at the effect of microstructure on the bifurcation of cracks in alumina polyurethane composites with herringbone structure [44]. They found that microstructures with platelet angles oriented at 60 ° were strongest and toughest when testing notched beams in the horizontal direction.

Our experimental results also show that monoliths with nacre-like horizontal alignment are the toughest. In multilayered composites, composites with 90 ° angle between each layer are also tougher than the other monoliths and the multilayers with lower angles. One motivation for creating the multilayered composites with twisting and tilting angles $\theta$ and $\Phi$ was to explore the effect of crack mixity on crack propagation and toughness. Indeed, these samples are reminiscent of the helicoidal, Bouligand-like composites where the increase in crack dissipation and toughness has been reported to result from the mode mixity during cracking [8,9,45]. Since analytic models have been derived based on orthotropic properties, we applied the same model on our microplatelet reinforced composites to assess the role of crack mixity in the fracture and toughness of our composites (see Supplementary Information for details about the model). According to the model, the toughness increases when platelet angles $\theta$ and $\Phi$ increase with respect to the vertical loading direction. $\theta$ and $\Phi$ of 90 ° therefore yield the highest energy dissipation and toughness. In our samples, 90 ° is indeed is the best and the matrix is weak enough to fail before the platelets. Indeed, in natural nacre, the maximum, critical angle at which platelets break instead of deflecting cracks is $\theta_c = 75°$ [46], and in nacre-like alumina, $\theta_c = 56°$ [47]. This suggests that the simple model captures the properties of our microplatelet reinforced matrix, whereas it would need to be adapted for ceramic composites where other mechanisms take place, such as mineral bridges and shear interlocking. In multilayered composites with varying $\theta$ and $\Phi$ of 90 °, the performance is decreased, presumably due to a weaker interface between the layers. Also, the layers with $\theta = 0°$ are not beneficial to the increase in toughness. Yet, it is noticeable that the ductility of the composites remained high.

**6| Conclusions**

Microplatelets orientation using magnetically-assisted slip casting could be controlled in 3D with tunable angles and layer heights to build microstructured porous scaffolds using a new setup. These scaffolds could be infiltrated with a polymeric matrix, PDMS, to form reinforced composites. Having the possibility to design the microstructure of composites, hybrid nacre-like helicoidal samples were fabricated to study the effect of microstructure on the mechanical properties. We therefore produced and tested a set of 8 microstructures, including monoliths and multilayers. Our results show that horizontal nacre-like microstructures are still the strongest and toughest under vertical compression loads, but that multilayers can reach higher stiffness while retaining some ductility. Thanks to the capability to carefully control the microstructure, this work could allow more careful studies on microstructure-properties relationships in bioinspired microstructured samples in view of creating materials with superior properties.

**Acknowledgements**

The authors acknowledge financial support from the National Research Foundation, Singapore, with the Fellowship NRFF12-2020-0006. We thank Xiangyu Zhou for measuring the change in color with respect to the orientation.

**Supporting information**

Supporting information is attached.

**Author contributions**

X.Y.C. designed the setup, prepared and characterised the samples. C.C. helped with the setup design and measured the slip casting kinetics. S.T. did the rheological characterization. H.L.F designed the study, did the analytical modelling, and wrote the manuscript. All authors read, discussed, and reviewed the manuscript.

**Conflict of interest**

The authors declare no conflict of interest.